\documentclass[aps,prb,twocolumn,amsmath,amssymb,gallery,footinbib,superscriptaddress,floatfix]{revtex4-2}

\usepackage{graphicx}
\usepackage{dcolumn}
\usepackage{bm}
\usepackage{hyperref}

\usepackage{xcolor} 

\begin{document}

\preprint{APS/123-QED}

\title{Berezinskii-Kosterlitz-Thouless phases in ultra-thin PbTiO$_3$/SrTiO$_3$ superlattices}

\author{Fernando G\'omez-Ortiz}
\affiliation{Departamento de Ciencias de la Tierra y F\'{\i}sica de la Materia Condensada, Universidad de Cantabria, Cantabria Campus Internacional, Avenida de los Castros s/n, 39005 Santander, Spain} 
\author{Pablo Garc\'{\i}a-Fern\'andez}
\affiliation{Departamento de Ciencias de la Tierra y F\'{\i}sica de la Materia Condensada, Universidad de Cantabria, Cantabria Campus Internacional, Avenida de los Castros s/n, 39005 Santander, Spain}
\author{Juan M. L\'opez}
\affiliation{Instituto de F\'{\i}sica de Cantabria (IFCA), CSIC - Universidad de Cantabria, 39005 Santander, Spain}%
\author{Javier Junquera}%
\email[Corresponding Author:~]{javier.junquera@unican.es}
\affiliation{Departamento de Ciencias de la Tierra y F\'{\i}sica de la Materia Condensada, Universidad de Cantabria, Cantabria Campus Internacional, Avenida de los Castros s/n, 39005 Santander, Spain}%

\date{\today}

\begin{abstract}
We study the emergence of Berezinskii-Kosterlitz-Thouless (BKT) phases in (PbTiO$_3$)$_3$/(SrTiO$_3$)$_3$ superlattices by means of second-principles simulations. 
Beyond a threshold tensile epitaxial strain of $\epsilon = 0.25 \%$ the local dipole moments within the superlattices are confined to the film-plane, and thus the polarization can be effectively considered as two-dimensional. 
The analysis of the decay of the dipole-dipole correlation with the distance, together with the study of the density of defects and its distribution as function of temperature, supports the existence of a BKT phase in a range of temperature mediating the ordered ferroelectric (stable at low $T$), and the disordered paraelectric phase that appears beyond a critical temperature $T_{\rm BKT}$.
This BKT phase is characterized by quasi-long-range order (whose signature is a power-law decay of the correlations with the distance), and the emergence of tightly bounded vortex-antivortex pairs whose density is determined by a thermal activation process.
The proposed PbTiO$_{3}$/SrTiO$_{3}$ superlattice model and the imposed mechanical boundary conditions are both experimentally feasible, opening the door for the first experimental observation of these new topological phases in ferroelectric materials. 
\end{abstract}
\maketitle
\section{Introduction}
\label{sec:introduction}

The existence or absence of long-range order in two-dimensional (2D) systems is a question that has attracted a lot of interest due to its fundamental relevance and potential for applications.
Almost ninety years ago, Peierls demonstrated that 2D crystalline long-range order is destroyed by the thermal motion of long-wavelengths phonons~\cite{Peierls-1935}.
This result was generalized by Mermin and Wagner for a one- or two-dimensional isotropic finite-range spin Heisenberg model~\cite{Mermin-1966} in a theorem that states that for a system of dimensions smaller or equal to two, with short-range interactions, there cannot be a spontaneous symmetry breaking of any continuous symmetry at finite temperature. 
Independently, Hohenberg ruled out the existence of long-range order in Bose and Fermi liquid systems and Cooper pairs in superconductors for one and two dimensions~\cite{Hohenberg-1967}.
In these systems, the existence of spontaneous, smooth, long-wavelength excitations prevents the formation of long-range order at any non-zero temperature.
All these findings were rigorously generalized in a mathematical theorem by Dobrushin and Shlosman~\cite{Dobrushin-75}.

However, the absence of long-range ordering in such systems does not exclude the existence of phase transitions associated with topological defects corresponding to singularities in the polarization field. 
The existence of topological phase transitions were beautifully exemplified by the 2D XY-model~\cite{Berezinskii-71,Berezinskii-1972,Kosterlitz73}, where spins at a given point in space can be oriented along any of the possible directions on the plane (i.e. they are rotors in the unit circle, $\mathbb{S}_{1}$).  In the absence of external magnetic fields, the Mermin-Wagner theorem ensures that the total magnetization is zero at any finite temperature, so that the system lacks long-range magnetic order. The system undergoes, however, a topological phase transition mediated by defects.  
Typical topological defects in the XY-model are vortices and antivortices, characterized by a non-vanishing winding number.
An isolated vortex (or antivortex) always displays an inhomogeneous distribution of the orientation of the rotors, even in regions far away from the singularity, and entails an energy cost that diverges logarithmically with the system size $L$~\cite{Kosterlitz73}. 
This energy is large enough to inhibit the formation of lone topological defects out of thermal fluctuations at low enough temperatures. Therefore, vortices or antivortices in isolated form do not appear. 
However, they can appear as tightly bound vortex-antivortex pairs.
Indeed, from a large-scale perspective, these pairs are topologically neutral:
the distortion produced by the vortex cancels out the distortion due to the antivortex in a sufficiently distant region and, therefore, orientational disturbance vanishes and the resulting texture can be immersed in a uniform background.
The energy cost of creating a vortex-antivortex pair does not diverge in the thermodynamic limit and only depends on the topological charge of both defects and the natural logarithm of the distance between their singularities~\cite{Kosterlitz73}.
For such a configuration of tightly bound vortex-antivortex isolated pairs, the spin-spin correlation function, even for infinitesimally small temperatures, decays as a power with the distance between the spins, $\langle \mathbf{S}(\mathbf{r})\cdot \mathbf{S}(\mathbf{0})\rangle \sim r^{-\eta(T)}$, with a temperature-dependent critical exponent $\eta(T)$~\cite{Berezinskii-71,Berezinskii-1972,Kosterlitz73}. Spin-waves prevent long-range magnetic order forcing the total magnetization to be zero, $\langle|\sum_{\mathbf{r}} \mathbf{S(\mathbf{r})}|\rangle = 0$, as the Mermin-Wagner theorem states for such systems.
Nevertheless, since the correlation decay is not exponential, the system is not in a fully disordered state either, but rather in a quasi long-range order state. 
As the temperature is raised, the number of pairs increases and larger ones start forming, screened by other smaller pairs that lie in between. The average separation between a vortex and an antivortex becomes comparable to the separation between pairs. These loose pairs become effectively unbound.
The origin of this phase transition can be traced back to entropic effects, which favor large orientational differences between neighboring spins.
Indeed, the entropy also diverges logarithmically with the system size, leading to a subtle logarithmic competition between the energy and entropy associated with defects through the free energy of the system, $F= E- TS$. For high enough temperatures, the energy cost of creating a defect can be compensated with the entropy gain of proliferating defects in the system, leading to a mimimum of the free energy.
At a critical temperature, coined as the Berezhinskii-Kosterlitz-Thouless (BKT) temperature, $T_{\rm BKT}$, the entropy gain exactly balances the interaction energy, allowing for single topological defects to wander in the system. 
As a result, a disordered phase appears above $T_{\rm BKT}$, in which the spin-spin correlation function decays exponentially with the dipole-dipole separation.

The physical mechanism described above explains the existence of topological phase transitions not only in the XY-model but can also be applied to a variety of 2D systems, including two-dimensional melting~\cite{Young-1979, Nelson-1979}, the physics of superfluid helium ~\cite{Bishop-1978}, superconductors~\cite{Beasley-1979,Hebard-1980,Wolf-1981}, Josephson junction arrays~\cite{Resnik-1981}, and nematic liquid crystals~\cite{Lammert-1993}, among others.

For a long time, it was believed that ferroelectrics could not support BKT phases as it was thought they violated the prerequisites for the appearance of tightly bound vortex-antivortex pairs for two main reasons.
First, the discrete symmetry: the ferroelectric polarization points along specific directions in space and typically is three-dimensional (3D).
Second, the presence of long-range dipole-dipole interactions, which tend to reduce fluctuations, favoring long-range order ferroelectric phases.
The situation changed dramatically merely five years ago, after the appearance of some theoretical predictions sustained on Monte Carlo simulations of a first-principles-based effective Hamiltonian, where an intermediate critical BKT phase, underlain by quasicontinuous symmetry, emerged between the ferroelectric and the disordered paraelectric phases in tensily strained thin films of BaTiO$_{3}$~\cite{Nahas-17}.
Due to tensile strain, local dipole moments are confined to the film-plane, and
thus polarization can be regarded as a 2D vector field. 
The fourfold anisotropy (the tendency of the polarization to point along the symmetry-equivalent $<110>$ directions) becomes relatively irrelevant in the temperature range where the BKT phase was found. 
Indeed, within this temperature range, a situation equivalent to the two-dimensional XY-model phenomenology could be attained and a quasicontinuous symmetry was observed.
However, this anisotropy is important at low temperatures, suppressing fluctuations and restoring the fourfold rotational symmetry.
In summary, short-range anisotropic and further-ranged dipolar interactions ineluctably drive ferroelectric long-range order at low temperatures.
This endows the system with a temperature-dependent three-phase structure: (i) a truly ordered ferroelectric phase,
(ii) a quasi-long-range ordered phase substantiated by
an algebraic decay of spatial correlations and supported by an emergent continuous symmetry that allows for stable topological defects to condense in the distortion-confining form of vortex-antivortex bound pairs, and 
(iii)  a disordered, paraelectric phase, with exponentially falling correlations at high-enough temperatures.
A subsequent simulation manuscript also suggested the presence of a BKT phase in another 2D ferroelectric material (one unit-cell-thick SnTe) under tensile epitaxial strain~\cite{Xu-20}. 

The previous works opened the research avenue for the quest of BKT-like phases in ferroelectric materials. Here we undertake the next step, predicting similar phases in PbTiO$_{3}$/SrTiO$_{3}$ superlattices, a system that has been widely grown with atomic level precision. 
This is a system that displays a rich phase diagram depending on periodicity, epitaxial strain, and electric boundary conditions. 
Indeed, depending on the lattice constant of the substrate and on the periodiciy of the superlattice, classical flux closure domains~\cite{Tang-15}, or continuously rotating polar vortices~\cite{Yadav-16}, skyrmions~\cite{Das-19}, merons~\cite{Wang-20} or supercrystals~\cite{Stoica-19} have been reported. 

In this Article we show how in short-period superlattices made of the repetition of three unit-cells of each material, under sufficiently large tensile strain conditions, a quasi-long-range ordered BKT phase appears separating an ordered ferroelectric phase, at low temperatures, and a disordered phase, at high enough temperature.
After presenting the methodology in Sec.~\ref{sec:methodology}, we discuss in Sec.~\ref{sec:poldist} how, beyond a threshold tensile epitaxial strain of 0.25 \%, the polarization can be effectively considered to become 2D. The dipole-dipole correlation functions, whose evolution with distance is a fingerprint of the ordered (finite value at infinite distances), quasi-long-range (power-law decay), or disordered (exponential decay to zero) states are computed in Sec.~\ref{sec:corre_fun}. The creation of tightly-bound vortex-antivortex pairs at the intermediate temperature, and their proliferation and unpinning at a critical temperature is the focus of Sec.~\ref{sec:unpinning}. The defect density that allows the estimation of the chemical potential for the formation of defects is computed in Sec.~\ref{sec:defect-density}. 
All simulations presented here support the existence of an intermediate BKT phase in PbTiO$_3$/SrTiO$_3$ superlattices under sufficiently large epitaxial tensile strain values.
The proposed PbTiO$_3$/SrTiO$_3$ superlattice model and the imposed mechanical boundary conditions are both experimentally feasible, which opens the door for the first experimental observation of these new topological phases in ferroelectric materials. 
\section{Methodology}
\label{sec:methodology}

We have carried out extensive second-principles simulations of (PbTiO$_{3}$)$_{3}$/(SrTiO$_{3}$)$_{3}$ superlattices under different tensile strain mechanical boundary conditions, and studied the effect of temperature on the ordering of the system. The calculations were done using  the  same  methodology  presented  in  previous works~\cite{Wojdel2013,Wojdel2014}, as implemented in the {\sc{scale-up}} package~\cite{Wojdel2013,Pablo2016}.
The second-principles parameters of both materials were fitted from density
functional theory imposing a hydrostatic pressure of $-11.2$ GPa to counter the
underestimation of the local density approximation.
Once this pressure is considered, the theoretical equilibrium lattice constant of the cubic phase was taken as the reference structure, with a value of $a=b= 3.901$ \AA~ forming an angle $\gamma= 90^{\circ}$. The different epitaxial tensile strains used during the simulations were computed with respect to that reference.
The interatomic potentials, and the  approach  to  simulate  the  interface,  are  the  same  as in Ref.~\cite{Shafer-18}. 
The lateral size, $L$, of the simulation supercell typically was $L=20~\rm{u.c}$, although larger supercells of $L=30,~40~\rm{u.c}$ were also used at specific temperatures in order to determine the finite-size scaling of the correlations.
For each tensile epitaxial strain, $\epsilon$, we solved the model by running Monte Carlo simulations at different fixed $T$, where the equilibrated final state at a given temperature was used as initial condition for the next increased temperature. This allowed to speed up considerably the simulations by starting from configurations that were closer to equilibrium than completely random samples. Each calculation comprises 20.000 thermalization sweeps followed by 20.000 sampling sweeps, where we saved the geometry of the supercell every 500 steps in order to compute averages. After collecting all the geometries we computed the dipole-dipole correlation functions. 
\section{results}
\label{sec:results}

\subsection{Polarization distribution as a function of the epitaxial strain}
\label{sec:poldist}

\begin{center}
  \begin{figure*}[!]
     \centering
      \includegraphics[width=\textwidth]{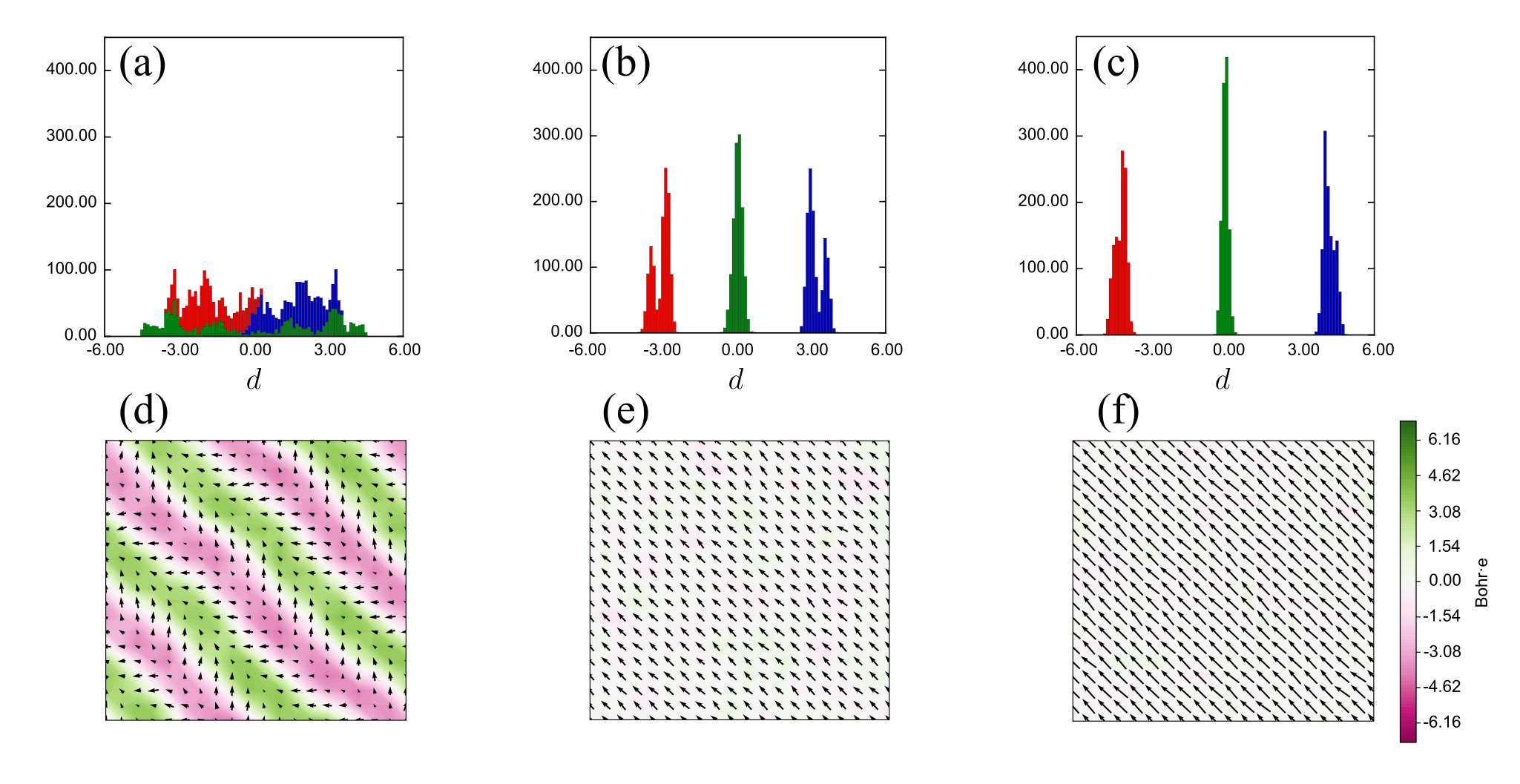}
      \caption{Polarization distribution as a function of the epitaxial strain.
      Distribution of the local dipole components at 25 K under a tensile epitaxial strain of 
      (a) 0.25 \%, (b) 1.00 \%, and (c) 3.00 \%.
      Red, blue and green histograms represent, respectively, the local dipole components 
      along [100], [010], and [001] pseudocubic directions. 
      The magnitude of the dipole moment $d$ is given in atomic units.
      Panels (d)-(f) represent the electric polarization textures, following the tensile epitaxial strain of the top row panels.
      Black arrows indicate the local dipoles, projected onto the $(x,y)$ plane. 
      The out-of-plane component of the polarization along the $[001]_{\rm pc}$ direction is represented by the green and magenta domains. 
      }
      \label{fig:order_param} 
  \end{figure*}
\end{center}

Epitaxial strain is a tuning parameter that has a strong influence on the order parameter space (i.e. on the possible values and orientation of the polarization at a given point in space) due to the strong polarization-strain coupling in ferroelectric oxide perovskites.
Previous first-principles results on (PbTiO$_{3}$)$_{3}$/(SrTiO$_{3}$)$_{3}$ superlattices support the stabilization of 180$^{\circ}$ stripe polydomain configurations, when the in-plane lattice constant of SrTiO$_{3}$ was imposed~\cite{Aguado-Puente-12}.  
Our second-principles results nicely reproduce this tendency for the low temperature ordered phase, that is maintained even for small tensile strains. As it can be seen in Fig.~\ref{fig:order_param}(a) for a tensile strain of 0.25\%,  the number of local dipoles with  a non-vanishing value of the polarization along the $[001]_{\rm pc}$ direction distributes symmetrically with respect to zero, spreading over relative large values, of the same order as the in-plane Cartesian components.   
Under this mechanical boundary condition, the order parameter space for the polarization is effectively 3D.
The stabilization of the 180$^{\circ}$ stripe domains in the out-of-plane direction for the low-temperature ordered phase is clearly visible in Fig.~\ref{fig:order_param}(d) coexisting with classical ferroelectric $a_{1}/a_{2}$ domains in the $(x,y)$ plane.

As the tensile strain is increased, the order parameter tends to be confined in the $(x,y)$ plane. For a 1\% value, the histogram corresponding to the $[001]_{\rm pc}$ Cartesian component shrinks, as shown in Fig.~\ref{fig:order_param}(b). This translates into the fact that almost all the out-of-plane components of the local dipoles are close to zero. Numerically the macroscopic polarization is found to lie along a $\langle110\rangle$ pseudocubic direction [the $[-110]$ direction in Fig.~\ref{fig:order_param}(b), but the energy is invariant with respect to any other symmetry equivalent direction].
Thus, the polarization can be effectively considered as a two-component vector.
Even more, the residual up (green) and down (magenta) domains in Fig.~\ref{fig:order_param}(e) are randomly distributed.
These effects are farther enhanced upon an increase of the tensile epitaxial strain up to 3\% [Figs.~\ref{fig:order_param}(c)-(f)]. 
\subsection{Dipole-dipole correlation function}\label{sec:corre_fun}
In order to study the effect of temperature on the ordering of the local polarization profile we computed the dipole-dipole correlation function, $C(r)=\langle \mathbf{P}(\mathbf{r})\cdot\mathbf{P}(\mathbf{0})\rangle$, where $\mathbf{r}=(r_x,r_y)$ and $r= \vert \mathbf{r} \vert$ represents the in-plane distance between dipoles.
The brackets, $\langle \dots\rangle$, stand for a triple average: first over all lattice site positions in a given $(x,y)$ plane; second, over the three PbTiO$_3$ planes in the simulation supercell along $z$; and third, over Monte-Carlo snapshots.

In an ordered system the polarization is aligned with a given direction and, therefore, the correlation function remains finite, even for infinite distances. 
Indeed, for an ordered system one expects to observe $\lim_{r \to \infty} C(r) \sim  R(T)$, where the asymptotic residual correlation value $R(T)$ is due to long-range ordering and, importantly, does not depend on system size.
As temperature is increased, random thermal fluctuations translate into a loss of long-range ordering and lower correlations at long distances, {\it i.e.} lower $R(T)$. Eventually, for high-enough temperatures, fluctuations are strong enough to destroy long-range polarization order, leading to a vanishing correlation function at large distances, $\lim_{r \to \infty} C(r) \sim f(T,L)$, with some system-specific L-decaying function $f$ such that, in the thermodynamic limit, $f(T,L \to \infty) = 0$. In both cases, the correlation is exponentially decaying at short distances, as corresponds to an interacting system, but the asymptotic limit behavior is clearly different for ordered and disordered phases.
This is the expected observation for the correlation function decay with distance in standard order/disorder transitions.

In contrast, if a BKT phase appears one would expect to have power-law decay of correlations, $C(r) \sim r^{-\eta(T)}$, with a temperature-dependent exponent below some critical temperature $T_{\rm BKT}$. BKT theory\cite{Kosterlitz73,Kosterlitz74} makes some precise predictions about the properties of the system at and below the transition. In particular, there is an universal critical exponent $\lim_{T \to T_{\rm BKT}^-} \eta(T) = 1/4$, which can be measured and used as a proxy for BKT-like behavior. For $T> T_{\rm BKT}$ the system shows large scale disorder, mediated by the mechanism of defect entropy gain described in the Introduction.

In this work we have studied this dependence of the correlation function with strain and temperature in our (PbTiO$_{3}$)$_{3}$/(SrTiO$_{3}$)$_{3}$ superlattices. Two different regimes can be distinguished.  

\subsubsection{Low tensile epitaxial strain: classical ferroelectric-paraelectric phase transition.}

\begin{center}
  \begin{figure*}[!]
     \centering
      \includegraphics[width=\textwidth]{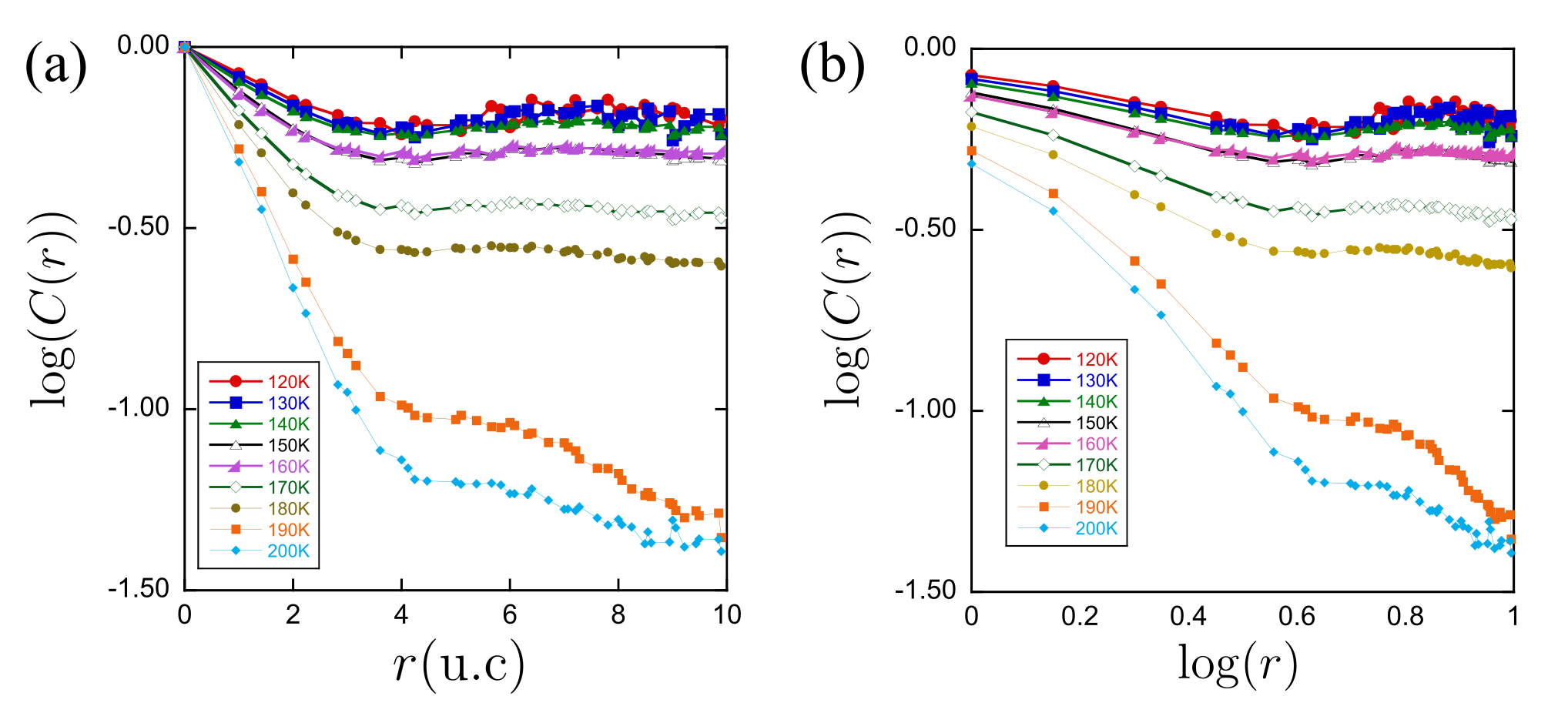}
      \caption{Two different representations of the correlation function as a function of the distance for a supercell of size $L=20$ under $\epsilon=0.25\%$ tensile epitaxial strain. (a) Logarithm of the correlation function against distance.
      (b) Logarithm of the correlation function against the logarithm of the distance. Symbols for different temperatures are shown in the inbox. u.c. stands for unit cell. 
      }
      \label{fig:low_tensile} 
  \end{figure*}
\end{center}

At low epitaxial strain values ($\epsilon\leq0.25\%$) the polarization along $z$ is comparable to that along in-plane directions ($x$, $y$), as shown in Fig.\ref{fig:order_param}(a) and Fig.\ref{fig:order_param}(d), making the polarization 3D. Under these conditions, with a 3D order parameter, BKT-like quasi-long range behavior cannot take place. If we observe the behavior of the logarithm of the correlation function against the distance upon increasing temperature [Fig.\ref{fig:low_tensile}(a)], two different regimes can be clearly identified. 
For $T<180$ K, the correlation falls rapidly down to a finite value and then stabilizes in a distance-independent plateau. This residual correlation decays with $T$, as corresponds to an long-range ordered state. 
In contrast, for $T>180$ K a rapid decay of the correlations is observed that tends to vanish for large distances. As explained before, we can ascribe this behavior to a disordered phase, where correlation rapidly falls following an exponential behavior, with shorter correlation lengths upon increasing temperature.
We can therefore identify the usual order-disorder (ferro-paraelectric) transition for low strain values. Log-log plots in Fig.~\ref{fig:low_tensile}(b) clearly show that a power-law decay for $C(r)$ is not compatible with the data for low tensile strains. 

\subsubsection{High tensile epitaxial strain: emergence of an intermediate BKT phase.}
\label{subSec:hightensile}
\begin{center}
  \begin{figure*}[!]
     \centering
      \includegraphics[width=\textwidth]{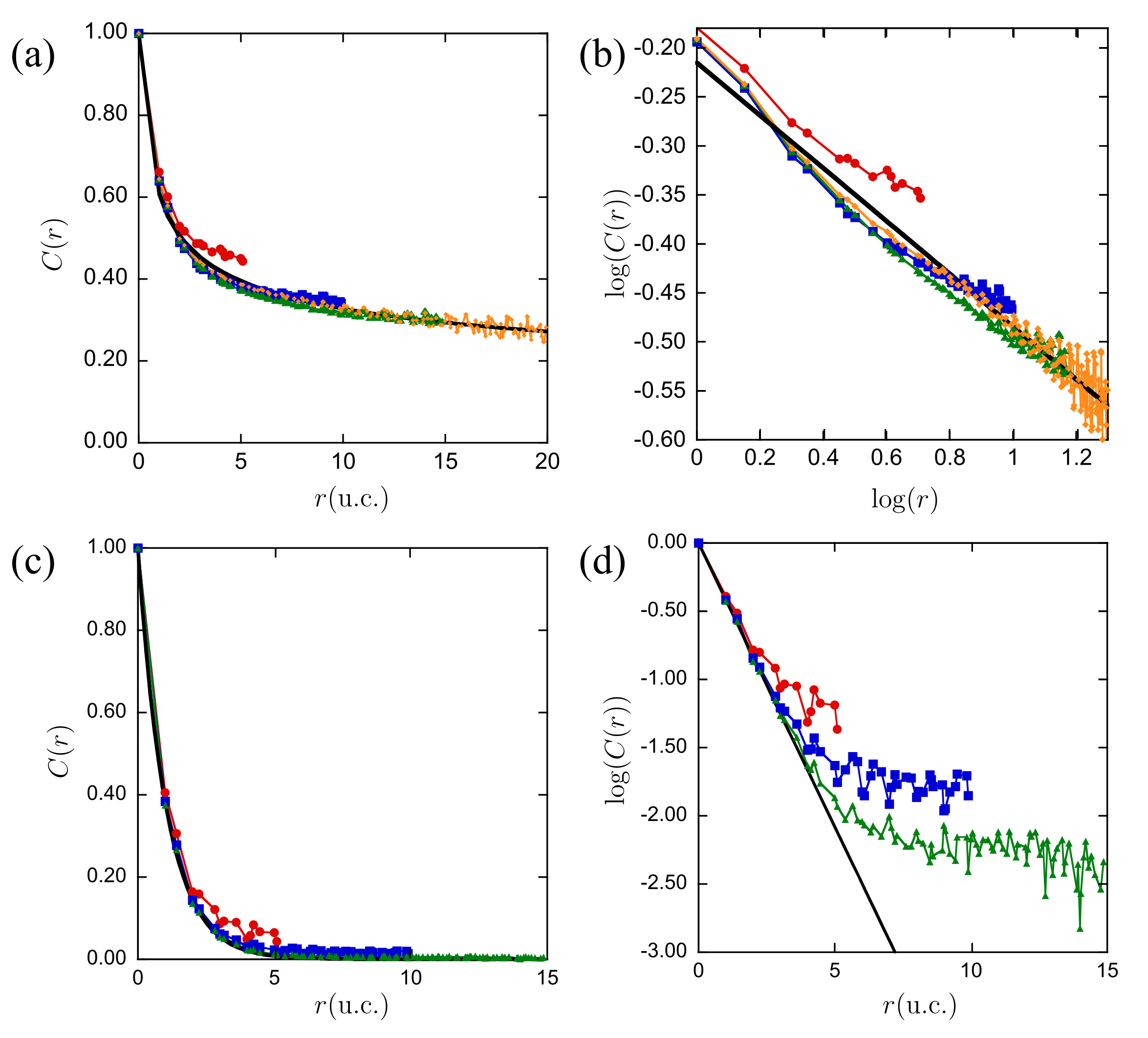}
      \caption{Correlation functions versus the dipole-dipole distance for different temperatures and sizes of the simulation box at constant epitaxial strain of $\epsilon=3\%$. 
      (a) $C(r)$ as a function of $r$ for $T = 430$ K. 
      (b) Same data, but in log-log scale. The same black solid line in panels (a) and (b) represents a linear fit of the data obtained for the largest simulation box. 
      (c) $C(r)$ as a function of $r$ for $T = 480$ K.
      (d) Same data, but taking the logarithm of the correlation function. 
      The same solid black line in panels (c) and (d) represents a linear fit for the data obtained at $L$ = 30 below a cutoff radii of 4.5 unit cells.
      Different lattice sizes $L=10,~20,~30,~40$ are represented by red squares, blue squares, green triangles and orange diamonds symbols, respectively.
      u.c. stands for unit cells.
    }
      \label{fig:size_scale} 
  \end{figure*}
\end{center}

Upon increasing epitaxial strain, the in-plane components of the polarization are favoured and, as shown in Fig.\ref{fig:order_param}, for high enough tensile strains the polarization can be considered effectively a 2D vector field. This dramatically affects the physics of the system and has a reflection in the behavior of dipole-dipole correlations as a function of distance, as shown in Fig.~\ref{fig:size_scale} for different temperatures and sizes of the simulation box.
Under a tensile epitaxial strain of $\epsilon=3\%$, and for a temperature of $T = 430$ K, the correlation function decays to zero with distance [Fig.~\ref{fig:size_scale}(a)]. This decay is not, however, exponential but power-law, as clearly evidenced by plotting the same data in a log-log plot, see Fig.~\ref{fig:size_scale}(b). Importantly, the power-law decay, $C(r) \sim r^{-\eta(T)}$, is independent of the system size, which is a strong indication of critical behavior. As we will see below, the critical region extends over a finite interval of temperatures with a temperature-dependent exponent $\eta(T)$. This is the expected phenomenology associated with BKT quasi long-range order. Further evidence showing that this is indeed a BKT phase will be presented in Sec.~\ref{sec:unpinning}, where we study the unbinding of vortex-antivortex pairs.

\begin{table}[b]
\caption{\label{tab:eta} Evolution of the critical exponent $\eta$ obtained as a linear fit in the $\log \left( C(r) \right)$ vs $\log(r)$ for the temperatures within the dashed oval of Fig.~\ref{fig:high_tensile} (b) for a supercell of size $L=20$ and tensile epitaxial strain of $\epsilon=1\%$. $R_\eta$ stands for the regression coefficient obtained in the fit to a power law decay, whereas $R_\xi$ stands for a regression coefficient obtained when fitting to an exponential decay. Temperatures in K. }
\begin{tabular}{ c  c  c  c}
\hline
\hline
T & $\eta(T)$& $R_{\eta}$ & $R_{\xi}$ \\ \hline
$240$ & $-0.040\pm0.003$ & $0.978$ & $0.690$\\
$245$ & $-0.072\pm 0.009$& $0.985$ & $0.709$\\
$250$ & $-0.127\pm 0.005$& $0.976$ & $0.769$\\ 
$255$ & $-0.27\pm 0.01$  & $0.977$ & $0.837$\\ 
$260$ & $-0.29\pm 0.02$  & $0.976$ & $0.838$\\
\hline
\end{tabular}
\end{table}

\begin{center}
  \begin{figure*}[hbtp]
     \centering
      \includegraphics[width=\textwidth]{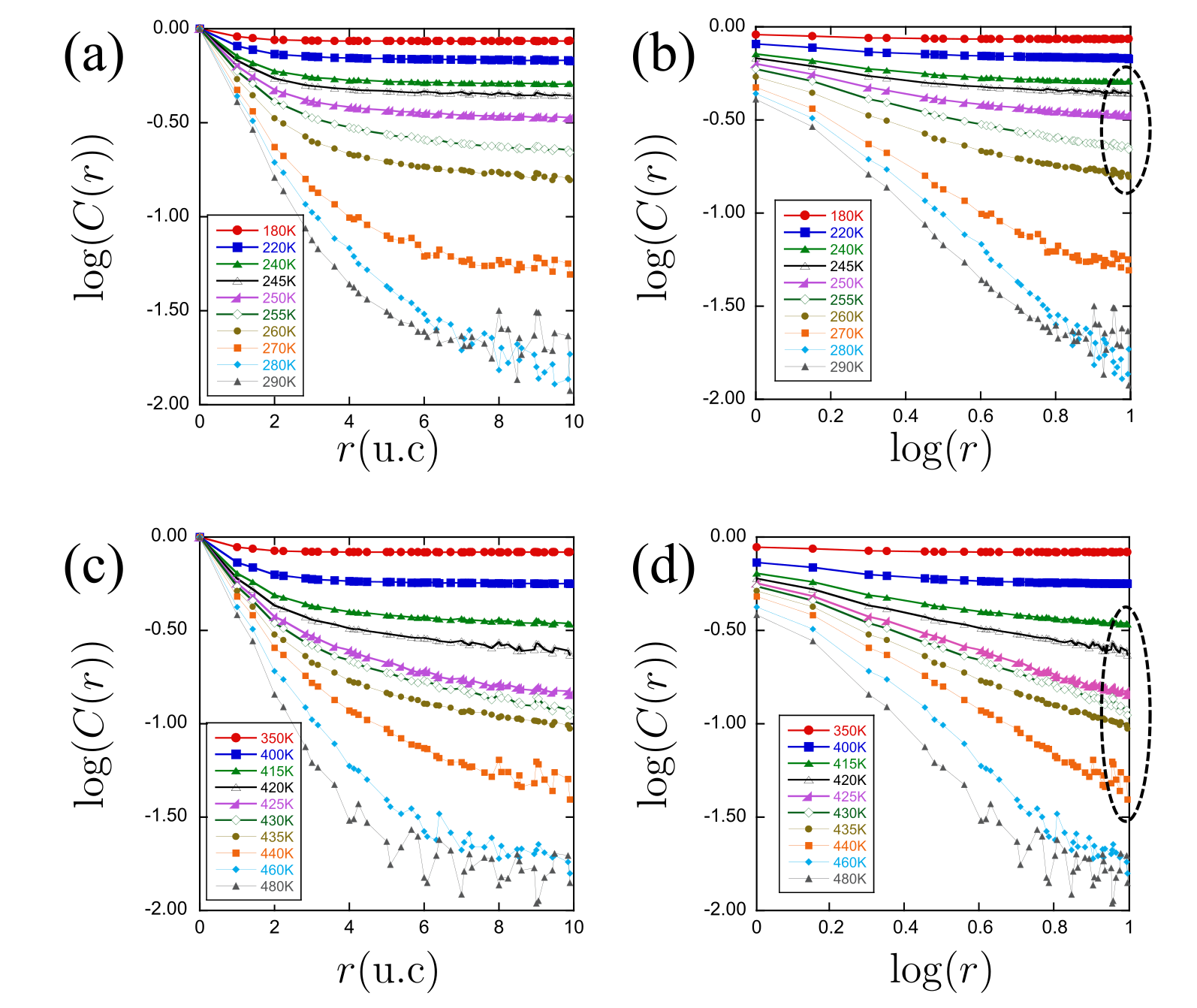}
      \caption{Two different representations of the dipole-dipole correlation as a function of the distance for a supercell of size $L$ =20 under different mechanical boundary conditions. (a)$\epsilon=1\%$. Logarithm of the correlation function against the distance. (b) $\epsilon=1\%$. Logarithm of the correlation function against the logarithm of the distance. (c) $\epsilon=3\%$. Logarithm of the correlation function against the distance. (d) $\epsilon=3\%$. Logarithm of the correlation function against the logarithm of the distance. Symbols for different temperatures are shown in the inbox. The dashed-oval at the right-hand side encapsulates the temperature range where the results from the simulations can be fitted to a straight line, with regression coefficient larger than 0.97.} 
      \label{fig:high_tensile} 
  \end{figure*}
\end{center}

Quasi long-range order is expected to be destroyed at high enough temperatures, due to the entropy increase associated with unbound vortex and antivortex freely wandering in the system. This is seen in Fig.~\ref{fig:size_scale}(c)-(d), where the dipole-dipole correlation becomes exponentially decaying for $T = 480$ K. Data in log-linear plot in Fig.~\ref{fig:size_scale}(d) clearly show that $C(r)$ goes to a residual correlation at large $r$ that decreases with the size $L$ of the simulation box, as corresponds to a disordered phase. 
In order to estimate the temperature range for the BKT quasi long-range order phase in the high tensile regime we study $C(r)$ vs. $r$ in Fig.~\ref{fig:high_tensile} (a)-(d) for $\epsilon = 1$\% and $\epsilon = 3$\% as temperature crosses the BKT critical point $T_{\rm BKT}$ for these two tensile strain samples.  
For high enough temperatures, thermal fluctuations destroy quasi long-range order and yield to the exponentially decaying correlations typical of a disordered system, as already discussed in Fig.~\ref{fig:size_scale}(d). 
This is evidenced by the linear behavior in the $\log C(r)$ vs $r$ plot in Fig.~\ref{fig:high_tensile}(b) (for temperatures above 260 K) and Fig.~\ref{fig:high_tensile}(d) (for temperatures above 440 K), in both high strain examples. 
In contrast, for low enough temperatures the correlation function at large distance tends to a plateau of residual correlation, typical of an ordered phase.
Remarkably, and in contrast with the low strain case, a critical region appears separating the ordered and disordered phases. This is the regime of interest, where correlations fall to zero with distance but its decay is a power-law rather than an exponential, as shown in Fig.~\ref{fig:high_tensile}(b) and  Fig.~\ref{fig:high_tensile}(d) for the range of temperatures within the oval.
As a comparison, see the log-linear and log-log plots in Fig.~\ref{fig:low_tensile}(a) and Fig.~\ref{fig:low_tensile}(b), respectively, clearly showing that a power-law decay for $C(r)$ is not compatible with the data for low tensile strains. 
For strain $\epsilon = 1$\% we find BKT power-law correlations within the range $T\in [240,260]$ K, while $T\in [415,440]$ K for $\epsilon = 3$\%. For temperatures above these intervals, the system is fully disordered (paraelectric) and it becomes ordered (ferroelectric) for temperatures below the range. Note that the upper bounds of these intervals correspond to the critical BKT temperature.
The values of the critical exponent $\eta(T)$ within the BKT phase can be obtained from a mean square-root fit to a straight line of $\log C(r)$ vs. $\log r$, as exemplified in Fig.~\ref{fig:size_scale}(b). The results of these fits for strain $\epsilon = 1$\% at different temperatures are summarized in Tab.\ref{tab:eta}.

All the results above agree with the picture expected when cooling a system from a classical disordered state to a critical BKT phase.
We note the crucial role played by the change of dimensionality of the polarization vector field, from 3D to 2D, induced by the mechanical tensile strain to obtain BKT behavior. 

We stress that, due to the intrinsic problems in studying BKT criticality (logarithmic convergence with the system size), it is very hard from a computational point of view to find the precise range of temperatures where this behavior is attained relying solely on the information of the correlation function and the fit of the critical exponent $\eta(T)$. In this sense the temperature intervals given should be considered as approximate. One would require much larger system sizes in order to precisely obtain the critical region, which is not computationally feasible within our second-principles simulations. 
Alternatively, one can obtain further evidence supporting the BKT phase picture in these range of temperatures by studying the density of defects and its distribution as a function of temperature. In the next section we focus on this issue and show that the results are consistently supporting BKT behavior.

\subsection{Unbinding and proliferation of defects}
\label{sec:unpinning}
As briefly discussed in the Introduction, there is a close relationship between the BKT transition and the disposition and behavior of polarization defects, {\it i.e.} vortices and antivortices.
As an additional piece of evidence for the existence of a BKT-like phase transition in our 
PbTiO$_{3}$/SrTiO$_{3}$ superlattices, we have computed the distribution of the vortex/antivortex pairs as a function of temperature.
To accomplish this goal, one first has to resolve the spatial positions where the vortices and antivortices are located.
This can be done calculating the topological charge at every lattice site by following the recipe explained in Appendix~\ref{app:a}.
At this point, it is important to note that with periodic boundary conditions, due to the Poincar\'e-Hopf theorem~\cite{Milnor65} (and energy conservation considerations), the total net vorticity shall always add to zero, {\it i.e.} the total charge of vortices present in the system at any given time must balance exactly with the charge of the antivortices.
This is the reason why one cannot observe an imbalance between vortices and antivortices throughout the system dynamics that should be, in principle, possible if the periodic boundary conditions could be relaxed.

\begin{center}
  \begin{figure*}[hbtp]
     \centering
      \includegraphics[width=\textwidth]{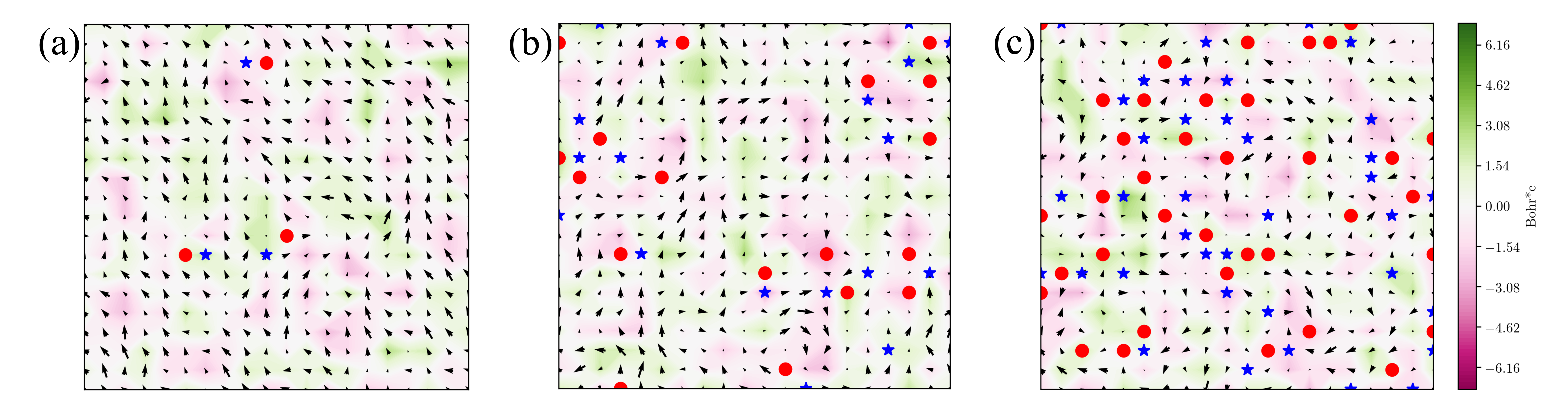}
      \caption{Distribution of topological defects for a superell of size $L=20$ and a tensile epitaxial strain of $\epsilon=1\%$ at different temperatures. (a) $T=220$ K, (b) $T\approx T_{\rm{BKT}}=255$ K and (c) $T=290$ K. Red dots represent vortices, with vorticity $n=+1$, whereas blue stars stand for antivortices of vorticity $n=-1$.}
      \label{fig:defects} 
  \end{figure*}
\end{center}

We show in Fig.~\ref{fig:defects} typical snapshots of the positions of the vortices and antivortices of the polarization pattern at three different temperatures for an epitaxial tensile strain of $\epsilon=1$\% after thermalization. 
In the low temperature regime, below $T_{\rm BKT}$ but within the BKT interval regime, $T\in [240,260]$ K, the typical quasi long-range order phenomenology is observed. A few tightly bound pairs of overall topologically neutral defects appear [Fig.~\ref{fig:defects} (a)], where the typical distance between the defect cores is one unit cell, which is the cheapest disturbance in energetic terms. Pairs of vortex-antivortex pop up at different positions of the lattice due to thermal fluctuations, wander around and annihilate randomly, while keeping the average density of pairs constant at a given temperature. From a large scale perspective, the orientational disturbance of a vortex/antivortex pair is canceled out.
But long-range order is destroyed by long-wavelength polarization excitations (equivalent to the spin-waves in the XY-model), which propagate in the system at zero energy cost,
leading to 
the quasi-long-range ordered state. On average, the total polarization remains zero in the BKT regime.
Upon increasing temperature the density of pairs increases [Fig.~\ref{fig:defects} (b)] until, above $T_{\rm BKT}$, the typical distance between pairs becomes comparable to the distance within pairs [Fig.~\ref{fig:defects} (c)]. At this point, thermal fluctuations dominate over vortex-antivortex pair interactions leading to configurations where neighboring vortex-vortex or antivortex-antivortex pairs, that should in principle be unfavorable, are observed. Vortex-antivortex pairs become effectively unbound.
The wandering of lone defects produces disorder in the system, making the pair correlation function to decay exponentially in agreement with the short-range ordering of the polarization pattern. 
\subsection{Defect density}
\label{sec:defect-density}
The importance of the entropic contribution translates in an increase of the density of defects $\rho$ as temperature is raised. This density is computed by averaging the number of vortex-antivortex pairs over both (i) the three different planes of PbTiO$_3$, and (ii) snapshots of the Monte Carlo simulation at a given temperature. The final number is divided by the supercell volume to get a density. 

\begin{center}
  \begin{figure}[!]
     \centering
      \includegraphics[width=\columnwidth]{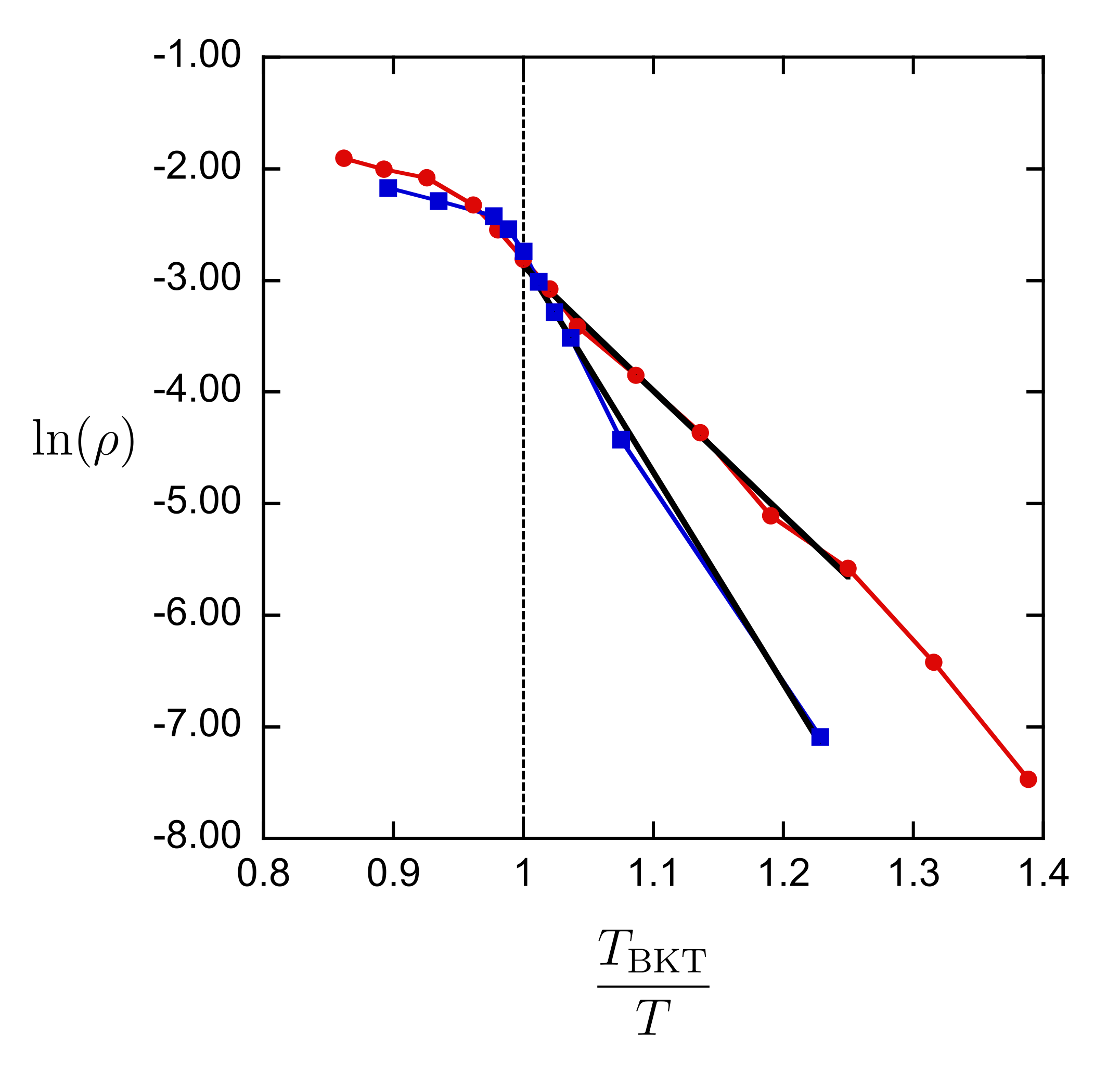}
      \caption{Density of vortex-antivortex pair defects as a function of temperature for a supercell of size $L=20$. Red dots and blue squares correspond to a tensile epitaxial strain of $\epsilon=1\%$ and $\epsilon=3\%$ respectively. Vertical dashed line is a guide to the eye for identifying the transition temperature at the point where the curvature changes. Black solid lines are a fit to a linear regression where the two last points are omitted in the case of $\epsilon=1\%$ as they correspond to 180 K and 190 K, where an ordered state is observed as shown in the steady plateau of Fig.~\ref{fig:high_tensile} (a).}
      \label{fig:density} 
  \end{figure}
\end{center}

Theoretically, at low temperatures, $T<T_{\rm BKT}$, one can make use of a dilute limit approximation and suppose that the density of pairs is so low than pair-pair interactions can be neglected. In this limit, the probability of exciting a vortex-antivortex pair is proportional to the Boltzmann factor $\exp[\frac{-\mu}{k_{\rm{B}}T}]$, where $\mu$ is the chemical potential to create a vortex-antivortex separated by a single unit distance. 
The numerical value of the chemical potential is specific of the particular system Hamiltonian. However, more importantly, the density of vortex-antivortex pairs is universally given by a thermal activation process. In the adequate units for energy one gets:
\begin{equation}
    \ln \rho(T) =-\alpha \left(\frac{T_{\rm BKT}}{T}\right) + {\rm const},
    \label{eq:density}
\end{equation}
where the additive constant comes from normalization. This equation is expected to hold below the critical temperature, within the BKT regime. Only as $T \to T^-_{\rm BKT}$ the chemical potential for creating vortices changes, due to the energy of pair unbinding, leading to deviations from Eq.~(\ref{eq:density}).
In Fig.~\ref{fig:density} we show the temperature dependence of the computed density of defects for high tensile strains and the fit to Eq.~(\ref{eq:density}).
The change in the curvature indicates the point where the dilute limit no longer holds. The deviation from Eq.~(\ref{eq:density}) serves as a fingerprint to detect the transition temperature, where sole defects start to proliferate provoking a change in the functional form of the chemical potential ($T\geq T_{\rm BKT}$).
From Fig.~\ref{fig:density}, we can infer a value of $T_{\rm BKT} \approx 255$ K for $\epsilon=1\%$ and $T_{\rm BKT} \approx 430$ K for $\epsilon=3\%$. Note that these values are consistent with our previous estimates [see Sec.~\ref{subSec:hightensile}], where we obtained a BKT regime of power-law decay of correlations that extends up to $T_{\rm BKT} \approx 260$ K and $T_{\rm BKT} \approx 440$ K respectively.

The defect density method to find the transition temperature is indeed more accurate for limited system-sizes than determinations based on the correlation function, which are prone to finite-size effects.
Actually, the precise numerical computation of the critical temperature is only attainable in the thermodynamic limit. Nevertheless, we observe that, when we define the transition temperature $T_{\rm BKT}$ from the fit to the defect density Eq.~(\ref{eq:density}) and compute the critical exponent of the power-law decay $C(r) \sim r^{-\eta(T)}$, the result nicely converges to something close to the universal value $\eta(T^-_{\rm BKT})=1/4$ theoretically predicted for BKT phase transitions. In Table.~\ref{tab:exponents} we summarize the values of the exponent $\eta(T_{\rm BKT})$ measured for a high tensile strain $\epsilon =3$\% at the defect density critical temperature in different lattice sizes ({\it i.e.} $L$ = 10, 20, 30, and 40, showing very good agreement with the theoretical prediction of the BKT theory. 
\begin{table}[b]
\caption{\label{tab:exponents}Evolution of the critical exponent $\eta$ obtained as a linear fit in the $\log(C(r))$ vs $\log(r)$ at $T_{\rm BKT}\approx430~K$ for $\epsilon=3\%$ as a function of the in plane lattice size. The critical temperature was obtained from the dependence of the density of defects with temperature given by Eq.~(\ref{eq:density}) and shown in Fig.~\ref{fig:density}}
\begin{tabular}{ c  c }
\hline
\hline
Lattice size & $\eta$ \\ \hline
$10\times10$ & $-0.23\pm0.01$ \\
$20\times20$ & $-0.240\pm 0.009$ \\
$30\times30$ & $-0.248\pm 0.005$ \\ 
$40\times40$ & $-0.247\pm 0.004$ \\ 
\hline
\end{tabular}
\end{table}
\section{Conclusions}
\label{sec:conclusions}

In summary, following the milestone works by Nahas and coworkers~\cite{Nahas-17} and Xu {\it et al.}~\cite{Xu-20}, we theoretically predict the existence of topologically non-trivial BKT phases in the experimentally well-studied PbTiO$_{3}$/SrTiO$_{3}$ ferroelectric/dielectric superlattices.
These phases only appear under sufficiently large tensile epitaxial strains, when the polarization becomes effectively 2D. 
At low-temperature, the long-range dipole-dipole interactions and short-range anisotropy favors a ferroelectric ordered phase.
Within this regime, the correlation function asymptotically tends to a system-size independent residual, whose value decreases with temperature. 
Upon increasing the temperature, the BKT phases are stabilized on a narrow, strain dependent, range of temperatures-- typically of the order of 20 K. 
They are characterized by the spontaneous thermally-activated formation of tightly bounded vortex-antivortex pairs of overall topologically neutral defects,
and by dipole-dipole correlation functions that decays as a power law with the distance, signature of quasi-long-range order. 
The temperature-dependent exponent of the power law tends to the critical universal value of 1/4 when approaching the critical temperature $T_{\rm BKT}$ from below.
Beyond this temperature, the vortex-antivortex pairs unbind and their density increases. The typical distance between pairs becomes comparable to the distance within pairs and the correlation function decays exponentially with the distance between dipoles. 

The presented results might have important consequences in other functional properties of the superlattices. In particular, the dielectric properties might depart from the classical Curie-Weiss behavior as a function of temperature expected for traditional ferroelectric to paraelectric phase transitions.
It is important to note that although the correlation function might not be experimentally accessible, the related electric susceptibility can be measured by standard techniques. 

The proposed PbTiO$_{3}$/SrTiO$_{3}$ superlattice model and the imposed mechanical boundary conditions are both experimentally feasible, opening the door for the first experimental observation of these new topological phases in ferroelectric materials.

\acknowledgments
F.G.-O., P.G.-F., and J.J. acknowledge financial support from Grant No. PGC2018-096955-B-C41 funded by MCIN/AEI/10.13039/501100011033 and by ERDF ``A way of making Europe'' by the European Union. 
F.G.-O. acknowledges financial support from Grant No. FPU18/04661 funded by MCIN/AEI/10.13039/501100011033. 
We thankfully acknowledge computing time at Altamira supercomputer and the technical support provided by the Instituto de F\'{\i}sica de Cantabria (IFCA) and Universidad de Cantabria (UC). We also thank J. \'A. Herrero for his valuable assistance with the supercomputing environment HPC/HTC cluster Calderon, supported by datacenter 3Mares from Universidad de Cantabria.
\appendix
\section{Calculation of the topological charge}
\label{app:a}
\begin{center}
  \begin{figure}[!]
     \centering
      \includegraphics[width=\columnwidth]{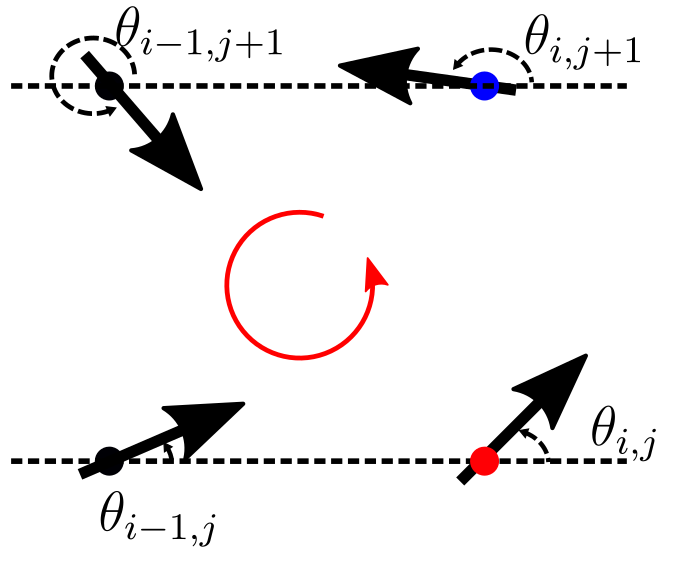}
      \caption{Sketch of a possible configuration of dipoles with vorticity $n=+1$ along an elementary plaquette of the simulation cell centered at the site $(i,j)$. The direction of the red arrow marks the sense of travel in order to compute the vorticity.}
      \label{fig:vorticity} 
  \end{figure}
\end{center}

The recipe followed to compute the topological charge is sketched in Fig.~\ref{fig:vorticity}.
For a given lattice site in the plane [marked with a red dot at the site $(i,j)$] we compute the angle between the local dipole and the horizontal axis ($\theta_{i,j}$ in Fig.~\ref{fig:vorticity}).
Then, we perform a closed counterclockwise loop over first neighbor sites in the $(x,y)$ plane [the first neighbour, at position $(i,j+1)$ is plotted with a blue dot in Fig.~\ref{fig:vorticity}]. 
For every neighbour site, we compute again the angle between the local dipole and the reference horizontal direction and subtract it from the one computed in the former position, mapping this difference in the $[-\pi,\pi]$ interval. 
The accumulated phase after closing the loop must be a multiple of $2\pi$. 
If it is zero, the vorticity at that particular lattice site vanishes.
Whenever the accumulated phase equals $2\pi$ or $-2\pi$, the corresponding site is associated with a vortex or an antivortex, respectively.
\end{document}